\documentclass[12pt]{article}
\usepackage{amssymb}

\renewcommand{\thefootnote}{\fnsymbol{footnote}}

\def\dspace{\baselineskip = 0.25in}

\begin{document}
\dspace
\begin{titlepage}
\begin{flushright}
BA-03-15
\end{flushright}
\vskip 2cm
\begin{center}
{\Large\bf TeV Scale Leptogenesis with Heavy Neutrinos}\\
\vspace*{1cm}
{\bf
S. Dar$^{\it a,}$\footnote[1]{sdar@udel.edu}, S. Huber$^{\it b,}$\footnote[2]{shuber@physik.uni-bielefeld.de}, V. N. {\c S}eno$\breve{\textrm{g}}$uz$^{\it a,}$\footnote[3]{nefer@udel.edu}\,and\,Q. Shaf\mbox{}i$^{\it a,}$\footnote[4]{shaf\mbox{}i@bxclu.bartol.udel.edu}}
\vskip 0.5cm
{\it $^a$ Bartol Research Institute, University of Delaware, \\Newark,
DE~~19716,~~USA\\[0.1truecm]}
\vskip 0.3cm
{\it $^b$ Universit\"at Bielefeld, Fakult\"at f\"ur Physik,
\\Universit\"atsstrasse. 25 33501 Bielefeld,~~Germany}

\end{center}
\vspace*{0.5cm}

\begin{abstract}
\noindent

Following a baryogenesis scenario proposed by Lazarides, Panagiotakopoulos
and Shaf\mbox{}i, we show how the observed baryon asymmetry can be explained via
resonant leptogenesis in a class of supersymmetric models with an intermediate mass
scale $M_I\lesssim 10^9$ GeV. 
It involves the out of equilibrium decay of heavy ($\lesssim M_I$) right handed neutrinos 
at a temperature close to the TeV supersymmetry breaking scale.
Such models can also resolve the MSSM $\mu$ problem.

PACS numbers: 12.60Jv, 11.30.Fs, 14.60.St, 98.80.Cq
\end{abstract}
\end{titlepage}
\renewcommand{\thefootnote}{\arabic{footnote}}
\setcounter{footnote}{0}
\setcounter{page}{1}

A large class of supersymmetric models possess {\it D} and {\it F} - f\mbox{}lat directions which can have important cosmological 
consequences. A particularly interesting set belongs to extensions of the minimal supersymmetric standard model (MSSM) and contains one or 
more intermediate to superheavy scales that arise from an interplay of a TeV scale from supersymmetry breaking and higher order 
(nonrenormalizable) terms suppressed by some cutoff scale $M_{*}$. Such models possess the following novel features that were discussed 
quite some time ago \cite{Lazarides:bj, s86, Lazarides:yq, npb93, mono,Yamamoto:1985rd,Binetruy:ss}, especially during the era of 
superstring inspired models:

\begin{enumerate}
\item In the context of the early universe the associated phase transition takes place at a temperature close to TeV, the supersymmetry breaking scale,  even though the gauge symmetry breaking scale is of intermediate size or higher \cite{Lazarides:bj, s86, Lazarides:yq};
\item The universe experiences a modest amount ($\sim 10$ or so e-foldings) of inf\mbox{}lation before the phase transition takes place \cite{s86, Lazarides:yq, npb93,Binetruy:ss}. This is now usually referred to as thermal inflation \cite{ti};
\item An appreciable amount of entropy generation occurs at the end of inf\mbox{}lation, and this could be exploited to dilute potentially troublesome relics such as super-heavy magnetic monopoles \cite{mono};
\item The f\mbox{}lip side of point (3) is that either a pre-existing baryon (or lepton) asymmetry should be sufficiently large to 
overcome the entropy onslaught, or a mechanism is in place to produce the asymmetry once the phase transition is completed. The latter case requires a final temperature of the radiation dominated Universe ($T_{f}$) in excess of an MeV (or so) to preserve hot big bang nucleosynthesis, and this sets an upper bound on the intermediate scale $M_I$ of around $10^{15}- 10^{16}~\textrm{GeV}$ \cite{Lazarides:yq}.
\end{enumerate}

In Refs.~\cite{s86, Lazarides:yq} a new mechanism for generating the baryon asymmetry was proposed, relying on the out-of-equilibrium 
decay of heavy (intermediate scale) particles at a temperature close to the TeV scale.
The novel feature here is that the decaying particles acquire mass through their coupling to the scalar field that is undergoing the phase 
transition. Thus, the phase transition and generation of baryon asymmetry occur more or less simultaneously. As noted in Ref. \cite{s86} 
the gravitino problem is neatly avoided in these models.

The main purpose of this paper is to show how the scenario of Refs. \cite{s86, Lazarides:yq} can be adapted to generate an initial 
lepton asymmetry, part of which is subsequently transformed to the observed baryon asymmetry \cite{ygd} through electroweak sphaleron mediated transitions \cite{spl}. We invoke resonant leptogenesis \cite{Flanz,Pilaftsis} to generate the required large initial asymmetry, before its dilution from entropy production.

The scenario we have in mind naturally arises in models based on subgroups of supersymmetric $SO(10)$ such as 
$H_1= SU(2)_{L}\times U(1)_{R}\times U(1)_{B-L}$ or $H_2=SU(2)_{L}\times SU(2)_{R}\times U(1)_{B-L}$ \cite{Pati:1974yy}. The Higgs field 
$\phi$, whose vacuum expectation value $\langle \phi\rangle\equiv M_I$ breaks $H_{1,2}$ to $SU(2)_{L}\times U(1)_{Y}$, should also 
provide masses comparable to $M_I$ to the right handed neutrinos. (For $H_2$, if $\phi$ belongs to the representation $(1,2)_{1}$, where 
the subscript labels the {\it B-L} charge, then dimension five operators will generate masses for the right handed neutrinos that are 
suppressed by the cutoff scale. However, if $\phi$ belongs to the representation $(1,3)_{2}$ of $H_2$, the right-handed neutrinos can acquire 
masses comparable to $M_I$. We will assume the latter case.)
The renormalizable part of the superpotential contains, among others, the following terms:
\begin{equation}\label{wr}
W_{R} \supset f_{ij} \phi N_i N_j+h_{i\alpha}N_{i}L_{\alpha}H_{u}\,,
\end{equation}
where $N_{i}$ denote the three right-handed neutrino superfields, $L_{\alpha}$ denote the three lepton superfields, $H_{u}$ is the 
MSSM doublet vacuum expectation value (VEV) that contributes to the neutrino Dirac mass and, unless otherwise stated, the 
dimensionless 
coefficients $f_{ij}$ are of order unity. The Yukawa couplings $h_{i\alpha}$ should be suitably chosen to reproduce the neutrino oscillation parameters.

In order to generate an intermediate scale VEV for $\phi$, the superpotential should not contain terms such as $\phi\overline{\phi}$ 
(the 
conjugate superfield $\overline{\phi}$ is present to ensure that supersymmetry is not broken at the intermediate scale). Furthermore, quartic terms consisting of the scalar component of $\phi$ and with coefficients of order unity must be absent from the potential. This is ensured by the gauge symmetry which forbids a cubic term for $\phi$ in the superpotential. The two most relevant dimension four (nonrenormalizable) terms in the superpotential are
\begin{equation}\label{wnr}
W_{NR}\supset \frac{\lambda}{M_*}(\phi\overline{\phi})^{2}+\frac{\beta}{M_{*}}\phi\overline{\phi} H_{u} H_{d}\,,
\end{equation}
where $M_{*}$ denotes the cutoff scale and $\lambda$ and $\beta$ are dimensionless coeff\mbox{}icients. The term proportional to 
$\beta$ is needed, as we will see, to ensure that the final temperature after completion of the phase transition is of order $10^{2} - 10^{3}$ GeV, so that the electroweak sphalerons can partially convert the lepton asymmetry to the observed baryon asymmetry.

The superpotential terms in Eqs.\,(\ref{wr}) and (\ref{wnr}) are easily realized by supplementing the gauge symmetries $H_{1,2}$ 
with suitable additional symmetries. For instance, in the $H_1$ case, a discrete symmetry $Z_2$ under which only $N_i$, 
$\overline{\phi}$, $H_d$ and $L_{\alpha}$ change sign is adequate. In the $H_2$ case, we can use a $Z_4$ symmetry with the following transformations: $(H_u,H_d)\to i\,(H_u,H_d)$, $N_i\to -i\,N_i$, $\phi\to-\phi$, with $\overline{\phi}, L_{\alpha}$ left unchanged. Such discrete symmetries may lead to the production of domain walls which, in principle, can be problematic. A resolution of the domain wall problem in this class of models has been extensively discussed in Ref. \cite{npb93}.

We see from Eq.\,(\ref{wnr}) that the combinations $H_{u}H_{d}$ and $\phi \bar{\phi}$ transform identically under any additional 
symmetries. Since $\phi \overline{\phi}$ is absent from the superpotential in order to generate a f\mbox{}lat direction, we are led to 
conclude that the `bare' MSSM $\mu$ term must also be absent. Thus, we have a nice mechanism for resolving the MSSM $\mu$ problem. The 
induced $\mu$ term $\beta M_{I}^{2}/M_{*}$ is of TeV scale as desired. (A resolution of the $\mu$ problem in this class of models has previously been discussed in Ref. \cite{Lazarides:bj}, as well as in the first paper in Ref. \cite{Dvali:1997uq}.)

Following common practice, we use $\phi$ to also denote the scalar component of the superfield $\phi$. Assuming that $\phi$ 
(sometimes referred to as a ``f\mbox{}laton'' \cite{Yamamoto:1985rd}) has sufficiently strong Yukawa couplings [Eq.\,(\ref {wr})] which 
can change the sign of its positive supersymmetry breaking mass squared term generated at some superheavy scale $\gg M_I$, and taking a 
D-flat direction where $\langle\phi\rangle=\langle\overline{\phi}\rangle^{\dagger}$, the zero-temperature effective potential of $\phi$ is 
\cite{s86, Lazarides:yq}
\begin{equation}\label{pot}
V_{0}(\phi)=\mu^4_0-M_{s}^{2}\big|\phi\big|^{2}+\frac{8\lambda^{2}}{M_{*}^{2}}\,\big|\phi\big|^{6}\,.
\end{equation} 
Here $\mu_{0}^{4}=(\frac{2}{3} M_{s}M_{I})^{2}$ is included to ensure that at the minimum $\langle \phi\rangle = M_{I}=(\lambda^{-1}M_{s}M_{*}/2 \sqrt{6})^{1/2}$, $V(M_{I})=0$, and $M_{s}$ ($\sim$ TeV) refers to the supersymmetry breaking scale.

For nonzero temperature the effective potential acquires an additional contribution, given by \cite{dj}
\begin{equation}\label{tpot}
V_{T}(\phi)=\left(\frac{T^{4}}{2\pi^{2}}\right)
\sum_{i} (-1)^{F}\int_{0}^{\infty}{\rm d}x\,x^{2}\,\textrm{ln}
\left(1-(-1)^F \textrm{exp}\{-\left[x^{2}+\frac{M_{i}^{2}(\phi)}{T^{2}}\right] \} \right),
\end{equation}
where the sum is over all helicity states, $(-1)^{F}$ is $\pm1$ for bosonic and fermionic states, respectively, and $M_{i}$ is the 
field-dependent mass of the {\it i}th state.
For $\phi\ll\textrm{T}$ Eq.\,(\ref{tpot}) yields a temperature-dependent mass term $\sigma \,\textrm{T}^{2}\big|\phi\big|^{2}$, where $\sigma\sim0.2$ for $f_{ij}\sim1$. Hence the potential
\begin{equation}\label{vt}
V(\phi)\,=\,\mu^4_0+(-M_{s}^{2}+\sigma \textrm{T}^{2})\big|\phi\big|^{2}+\frac{8 \lambda^{2}}{M_{*}^{2}}\,\big|\phi\big|^{6}
\end{equation}
has a minimum $V(\phi)\,=\,\mu^4_0$ at $\phi=0$ for $T\,>\,T_{c}= M_{s}/\sigma^{1/2}$. For  $\phi>T$, the temperature-dependent term is exponentially suppressed and $V(\phi)$ develops another minimum at $\phi=M_{I}$ for $T\,\lesssim \,M_{I}$.
 $\phi=0$ is the absolute minimum for $\mu_0 \lesssim T \lesssim M_{I}$ since the symmetric phase ($\phi=0$) has more massless degrees of freedom and the radiation energy density dominates over the false vacuum energy density $\mu^4_0$.
For $T\lesssim \mu_0$ the broken phase ($\phi=M_{I}$) becomes the absolute minimum of the potential, with $V(M_I)=0$. [The recently 
measured vacuum energy density of order ($10^{-3}$ eV)$^4$ is negligible for our purposes.]

The universe remains at $\phi=0$ for $T>T_{c}$ and, for $M_I\sim10^8$ GeV, experiences roughly $\ln(\mu_0/T_c)\sim 6$ {\it e}-foldings 
of inf\mbox{}lation due to the false vacuum energy density $\mu_0^4$ \cite{s86, Lazarides:yq, npb93}. During this phase the right-handed 
neutrinos $N_{i}$ are in thermal abundance
\begin{equation}\label{la}
\frac{n_{N_{i}}}{s}\,\simeq\,\frac{n^{eq}_{N_{i}}}{s}\,=\,\frac{45 \zeta(3)}{2 \pi^{4}}\,\frac{1}{g_{*}}\,\big(\frac{3}{4} 
g_{N_{i}}+g_{\widetilde{N}_{i}}\big)\,\simeq\,\frac{1}{300}\,,
\end{equation}
where $g_*$ counts the effectively massless degrees of freedom and $(g_{\widetilde{N}_{i}})g_{N_{i}}$ counts the degrees of freedom of 
({\it s})neutrinos.
When the temperature reaches $T_{c}$, the minimum (and the associated barrier) at $\phi=0$ disappears, and $\phi$ starts to roll down towards the minimum at $\phi=M_{I}$. The classical evolution of $\phi$ field is governed by the equation
\begin{equation}\label{dotphi}
\ddot{\phi}\,+\,3H\dot{\phi}\,=\,-\frac{dV}{d\phi}\,.
\end{equation} 
For $T<\phi<M_{I}$ the temperature-dependent mass term and the term proportional to $\big|\phi\big|^{6}$ can be ignored. Also, with the Hubble constant $H=\mu_0^{2}/\sqrt{3}M_{P}\ll M_{s}$ (where $M_P=2.4\times10^{18}$ GeV is the reduced Planck mass), Eq. (\ref{dotphi}) yields
\begin{equation} 
\ddot{\phi}\,\simeq\,M_{s}^{2}\,\phi\,, 
\end{equation} 
so that
\begin{equation}\label{phi_t}
\phi(\delta t)\simeq\,T_{c}\,\textrm{exp}(M_{s}\,\delta t)\,.
\end{equation}
From Eq.\,(\ref{phi_t}), it takes the f\mbox{}laton $\delta t\simeq\ln[M_I/T_c]/M_s \sim 10 M_{s}^{-1}$ to roll down to its minimum at $M_I$ \cite{Lazarides:yq, npb93}.

As the f\mbox{}laton rolls down, the right-handed neutrinos pick up a mass proportional to $\langle \phi \rangle$ and can decay out of 
equilibrium via the couplings $h_{i\alpha}N_{i}L_{\alpha}H_{u}$. The decay width is $\Gamma_{N_i}\simeq 
\sum_{\alpha}|h_{i\alpha}|^2\,M_{N_i}/8\pi$, where $M_{N_i}$ is the mass of the right-handed neutrinos when they decay. 

We show later that the resulting lepton asymmetry can account for the present baryon asymmetry of the universe for $M_s\ll M_I\lesssim(M_s/{\rm TeV})\times10^8$ GeV. (We will assume throughout that $M_I$ and $M_{N_i}\gg M_s$, otherwise thermal effects and direct f\mbox{}laton decay into neutrinos would modify our discussion.)
Since $M_{N_i}\lesssim M_I$, the light neutrino masses ($m_{\nu_{i}} \lesssim 0.1$ eV \cite{wmap}) require that the Yukawa couplings 
$h_{i\alpha}\lesssim (M_s/{\rm TeV})^{1/2}\times10^{-3}$, so that the decay time of the right-handed neutrinos $\Gamma^{-1}_{N_i}\gtrsim 
({\rm TeV}/M_s)\times10^3\,M_{s}^{-1}$. That is, they decay after the f\mbox{}laton has reached its minimum [but still long before the 
f\mbox{}laton decays, see Eq.\,(\ref{gammaphi})]. With $f_{ij}\sim1$ in Eq.\,(\ref{wr}), the mass of the right-handed neutrinos when 
they decay is $M_{N_i}\sim M_I$. (The assumption $f_{ij}\sim1$ can be relaxed without changing the main conclusions of this paper. For 
instance, we could have the third family right-handed neutrino mass $M_{N_3}\sim M_I$, whereas the first two family neutrinos are lighter.)

To be able to generate a lepton asymmetry, we must ensure that the right-handed neutrinos do not annihilate before they have time to 
decay. The annihilation rate for $N_{i}$ through $B-L$ gauge interactions is $\Gamma_{a}\sim \sum_{j}|f_{ij}|^2 \, T^{3}/8 \pi\langle\phi\rangle^{2}$. For $T \simeq T_{c}$, we estimate the annihilation probability $P_{a}$ before the flaton reaches the minimum to be $P_{a}=1-\textrm{exp}[\int_{0}^{\delta t}\Gamma_a \,{\rm d}t]\simeq \sum_{j}|f_{ij}|^2/(120\,\sigma^{1/2})\sim1/50$. Hence the number densities of $N_{i}$ do not change significantly before they decay, at least not from this process. Similarly, the annihilation of $N_{i}$ via dimension five couplings is also negligible. We therefore conclude that the $N_{i}$ do not annihilate before they have time to decay.  

The initial lepton asymmetry created by the decay of $N_{i}$ is diluted by entropy production and also partially converted to the baryon asymmetry \cite{ygd} by the sphaleron transition \cite{spl}. From the observed baryon asymmetry $n_{B}/s\simeq 8.7\,\times\,10^{-11}$ \cite{wmap}, the final lepton asymmetry is required to be $n_{L}/s\,\simeq\, 2.4\,\times\,10^{-10}$.
To see how much initial lepton asymmetry is needed to account for this value, we first estimate the final temperature $T_f$ and the dilution factor $\Delta$.

The f\mbox{}laton, with mass $m_{\phi}=2 \sqrt{2}M_s$ mainly decays via the superpotential coupling $(\beta/M_{*}) \phi \bar{\phi} H_u 
H_d$. (Recall that the $\mu$ parameter, also induced by this term in the superpotential, is naturally of order $M_s$ [$\mu\sim \beta 
M_I^{2}/M_{*}\sim(\beta/\lambda) M_{s}$]). The decay width of the f\mbox{}laton is
\begin{equation}\label{gammaphi}
\Gamma_{\phi} \simeq \frac{\beta^{2}}{8 \pi}\frac{M_{I}^{2}}{M_{*}^{2}}\, m_{\phi}\,=\,\frac{1}{24 \sqrt{2} \pi}\frac{\beta^2}{\lambda^{2}}\, \frac{M_s^{3}}{M_{I}^{2}}\,,
\end{equation}
so that $\tau_{\phi}=\Gamma^{-1}_{\phi} \sim (M_{I}/M_{s})^{2}(\lambda^{2}/\beta^{2})\,M_{s}^{-1}\gg M_{s}^{-1}$. For the final temperature $T_{f}\simeq\,0.3 (\Gamma_{\phi} M_{P})^{1/2}$, we find
\begin{equation}\label{reheat}
T_{f}\simeq\left[\frac{\beta}{\lambda}\left(\frac{M_s}{\rm{TeV}}\right)^{\frac{3}{2}}
\left(\frac{10^8~\rm{GeV}}{M_I}\right)\right]\times15~{\rm TeV}\sim M_{s}\,\textrm{for}\,\beta \sim 0.1\,.
\end{equation}
Note that the f\mbox{}laton decay products acquire  a plasma mass $\sim gT$ \cite{Weldon:bn} where $g$ is the $B-L$ gauge coupling. The f\mbox{}laton decay thus can only take place once the temperature drops below $\sim m_{\phi}/g$. Consequently the final temperature $T_f$ remains below $\sim m_{\phi}/g$ even for $M_I\ll10^8$ GeV. We will assume for simplicity that the final temperature $T_f\sim M_s$.

It is gratifying that $T_{f}$ is in a range where the electroweak sphalerons are able to convert some fraction of the lepton asymmetry 
into baryon asymmetry. This could not have been accomplished without the non-renormalizable term proportional to $\beta$ in 
Eq.\,(\ref{wnr}). [Integrating out $N$ from Eq.\,(\ref{wr}) yields an effective dimension six operator which gives a $\phi$ decay rate $\Gamma\sim M^5_s/M^4_I$, and the final temperature with only this decay would be of order GeV.] With $M_s\sim$ a few TeV, $\beta\sim0.1$ leads to a $\mu$ term in the range of a few hundred GeV, as desired.

The entropy production due to $\phi$ decay dilutes the initial lepton asymmetry by a factor
\begin{equation}
\Delta\,\simeq\,\frac{4 \mu_0^{4}/3 T_{f}}{(2 \pi^{2}/45) g_{*}(T_{c})T_{c}^{3}}\,\simeq\, \frac{3 \mu_0^{4}}{g_{*}T_{c}^{3}T_{f}}\,,
\end{equation} 
where $g_*=228.75$ for MSSM. Expressing the false vacuum energy density $\mu^4_0$ and the critical temperature $T_c$ in terms of $M_s$ and $M_I$ we obtain
\begin{equation}\label{dill}
\Delta\simeq5\times10^{-3}\,\frac{\sigma^{3/2}M_I^2}{M_s\,T_f}\,.
\end{equation}
We should make sure that the lepton asymmetry generated initially is large enough to sustain the impact of $\Delta$. The lepton asymmetry after dilution is given by 
\begin{equation}\label{las}
\frac{n_{L}}{s}\,=\,\sum_i\frac{n_{N_i}}{s}\,\frac{1}{\Delta}\,\epsilon_i\,.
\end{equation}
Here $\epsilon_i$ is the lepton asymmetry produced per decay of the $i$'th family neutrino $N_i$. 
Using Eqs.\,(\ref {la}), (\ref{dill}) and $T_{f}\sim M_{s}$  we get
\begin{equation}\label{la1}
\frac{n_{L}}{s}\,\sim\,\sum_i 5\,\left(\frac{0.2}{\sigma}\right)^{3/2}\left(\frac{M_{s}}{M_{I}}\right)^2\,\epsilon_i\,.
\end{equation}
For nearly degenerate neutrinos $\epsilon_i$ is given by \cite{Flanz,Pilaftsis}
\begin{equation}\label{cp1}
\epsilon_{i}\,\simeq\,\sum_{j\ne i} \frac{\textrm{Im}(h_{i\alpha}^{*}h_{\alpha 
j})^{2}}{\big|h_{i\alpha}\big|^{2}\big|h_{j\alpha}\big|^{2}}~\frac{\Delta M_{N}^{2} M_{N_{i}}\Gamma_{N_{j}}}{(\Delta M_{N}^{2})^{2}+ M_{N_{i}}^{2} \Gamma_{N_{j}}^{2}},
\end{equation}
where $\Delta M_{N}^{2}=M_{N_{i}}^{2}-M_{N_{j}}^{2}\simeq 2M_{N_i}(M_{N_{i}}-M_{N_{j}})$. Defining $\xi_{ij}=(M_{N_i}-M_{N_j})/(\Gamma_{N_j}/2)$, we can rewrite Eq.\,(\ref{cp1}) as
\begin{equation}\label{cp2}
\epsilon_i\,\simeq\,\sum_{j\ne i} \frac{\textrm{Im}(h_{i\alpha}^{*}h_{\alpha j})^{2}}{\big|h_{i\alpha}\big|^{2}\big|h_{j\alpha}\big|^{2}}~\frac{\xi_{ij}}{\xi^2_{ij}+1}\,.
\end{equation}
Assuming that $\delta_{CP}\equiv\textrm{Im}(h_{i\alpha}^{*}h_{\alpha j})^{2}/(\big|h_{i\alpha}\big|^{2}\big|h_{j\alpha}\big|^{2})$ is of order unity, the final lepton asymmetry is given by

\begin{equation}\label{lepas}
\frac{n_{L}}{s}\sim\sum_{i,\,j\ne 
i}\,2.4\times10^{-10}\,\left(\frac{0.2}{\sigma}\right)^{3/2}\left(\frac{M_{s}}{\textrm{TeV}}\right)^2\left(\frac{10^{8}\,\textrm{GeV}}{M_{I}}\right)^2\frac{\xi_{ij}}{\xi^2_{ij}+1}\,.
\end{equation}
The asymmetry is maximized when the resonance condition $\xi_{ij}=1$ is satisfied for at least one pair of families. This gives an 
upper bound on the intermediate scale $M_I\sim(M_s/{\rm TeV})\times10^{8}$ GeV, which corresponds to a cutoff scale $M_*\sim(M_s/{\rm TeV})\times10^{14}$ GeV for $\lambda\sim1$. With somewhat larger values for $M_s$, say about 10 TeV, the upper bounds are $M_I\sim10^9$ GeV and $M_*\sim10^{15}$ GeV. 
One could ask how this relatively low cutoff scale
 can be incorporated within a more fundamental theory. One
possibility is related to superstring inspired models with intermediate
cutoff scales which have been of much recent interest. Another possibility
is to introduce intermediate mass scale particles whose exchange can generate
an effective cutoff scale of the desired magnitude, even though the underlying
theory may have a cutoff scale that is significantly higher.

For $M_I\lesssim(M_s/{\rm TeV})\times10^{8}$ GeV the resonance condition does not have to be satisfied, although nearly degenerate right 
handed neutrinos are still needed. Suppose that the neutrino mass differences $M_{N_{i}}-M_{N_{j}}$ are much greater then the decay 
widths $\Gamma_{N_j}$ ($\xi_{ij}\gg1$), so that $\epsilon_i\sim\sum_{j\ne i}\xi^{-1}_{ij}$. [Equation (\ref{cp2}) in this case reduces 
to 
the perturbative result \cite{Covi:1996wh}.] Using the seesaw relation $\sum_{\alpha}\big|h_{j\alpha}\big|^{2}\sim (m_{\nu_j}/\langle 
H_u\rangle^2)\,M_{N_{j}}$ with $M_{N_{j}}\sim f_{j}M_{I}$ (where $f_j$ denotes an eigenvalue of $f_{ij}$), we can write
\begin{equation}\label{gamma}
\Gamma_{N_j}=\sum_{\alpha}\frac{\big|h_{j\alpha}\big|^2 M_{N_j}}{8\pi}\sim \frac{m_{\nu_j}\,f_{j}^{2}\,M^2_I}{8\pi\,\langle 
H_u\rangle^2}\left(\frac{m_{\nu_j}}{0.1{\rm eV}}\right)\left(\frac{M_I}{10^8~\rm{GeV}}\right)^2\times~f_{j}^{2}\,{\rm GeV}.
\end{equation}
Substituting Eq.\,(\ref{gamma}) in Eq.\,(\ref{lepas}), we obtain
\begin{equation}
\frac{n_{L}}{s}\sim\sum_{i,\,j\ne 
i}\,2.4\times10^{-10}\,\left(\frac{0.2}{\sigma}\right)^{3/2}\left(\frac{M_{s}}{\textrm{TeV}}\right)^2\left(\frac{m_{\nu_j}}{0.1~{\rm eV}}\right)\left(\frac{f_{j}^{2}\,{\rm GeV}}{M_{N_i}-M_{N_j}}\right)\,. 
\end{equation} 
The final lepton asymmetry is thus consistent with the observed baryon asymmetry provided that $M_s\ll M_I\lesssim(M_s/{\rm 
TeV})\times10^{8}$ GeV and that at least one pair of right-handed neutrino families have a mass difference less than or of order a GeV.

In conclusion, following Refs. \cite{s86, Lazarides:yq}, we have shown that in what are often referred to as thermal inflation models, 
there exists a novel mechanism for explaining the observed baryon asymmetry via leptogenesis. Because of significant entropy production 
that follows thermal inflation, the lepton asymmetry initially produced by heavy right-handed neutrinos with masses less than or of 
order $M_I$ (but greater than the flaton mass) must be as large as possible. This requires nearly degenerate right-handed neutrinos with 
GeV scale mass differences. It remains to be seen how this degeneracy can be realized in conjunction with realistic neutrino masses and 
mixings. To ensure that the electroweak sphalerons can partially convert the lepton asymmetry to the observed baryon asymmetry, we require 
that the final temperature after completion of the phase transition is of order $10^3$ GeV. This leads to the introduction of a term 
in the superpotential [Eq.\,(\ref{wnr})] which is also key to the resolution of the MSSM $\mu$ problem. Finally, it is clear that for 
intermediate scales significantly above $10^{9}$ GeV, leptogenesis should arise from the decay products of the flaton field. 
For baryogenesis this has been discussed in Ref. \cite{Lazarides:yq}.
\section*{Acknowledgments}
This work was supported by the U.S. DOE under Contract No. DE-FG02-91ER40626.


\end{document}